\documentclass[a4paper,english,aps,pra,twocolumn]{revtex4-1}
\usepackage[latin9]{inputenc}
\usepackage{float}
\usepackage{amsmath}
\usepackage{amssymb}
\usepackage{graphicx}

\makeatletter

\usepackage{color}
\usepackage{array}
\usepackage{textcomp}
\usepackage{bm}
\usepackage{multirow}
\usepackage{amssymb}
\usepackage{hyperref}
\hypersetup{colorlinks=true, citecolor=blue, urlcolor=blue, linkcolor=blue}
\usepackage{upgreek}
\usepackage{braket}

\makeatother

\usepackage{babel}
\begin{document}
\title{Echoes in a Single Quantum Kerr-nonlinear Oscillator}
\author{I. Tutunnikov,}
\thanks{Corresponding author: ilia.tutunnikov@weizmann.ac.il}
\affiliation{AMOS and Department of Chemical and Biological Physics, The Weizmann
Institute of Science, Rehovot 7610001, Israel}
\author{K. V. Rajitha, and I. Sh. Averbukh}
\affiliation{AMOS and Department of Chemical and Biological Physics, The Weizmann
Institute of Science, Rehovot 7610001, Israel}
\begin{abstract}
Quantum Kerr-nonlinear oscillator is a paradigmatic model in cavity
and circuit quantum electrodynamics, and quantum optomechanics. We
theoretically study the echo phenomenon in a single impulsively excited
(``kicked'') Kerr-nonlinear oscillator. We reveal two types of echoes,
``quantum'' and ``classical'' ones, emerging on the long and short
time-scales, respectively. The mechanisms of the echoes are discussed,
and their sensitivity to dissipation is considered. These echoes may
be useful for studying decoherence processes in a number of systems
related to quantum information processing.
\end{abstract}
\maketitle

\section{Introduction \label{sec:Introduction}}

Echoes in physics can be defined as spontaneous delayed responses
following a series of pulsed excitations. Perhaps the most famous
example is the spin echo effect \citep{Hahn1950PR,Hahn1953} conceived
by E. Hahn in 1950. The effect is induced by irradiating a collection
of spins by two delayed magnetic field pulses, resulting in a magnetization
response appearing at twice the delay between the pulses. Over the
years, echoes have been discovered in various physical systems, such
as systems consisting of many interacting/non-interacting particles,
or single quantum particles. Examples of echoes in many-particle systems
include photon echoes \citep{Kurnit1964,Mukamel1995}, neutron spin
echo \citep{Mezei1972}, cyclotron echoes \citep{Hill1965}, plasma-wave
echoes \citep{Gould1967}, cold atom echoes in optical traps \citep{Bulatov1998,Buchkremer2000,Herrera2012},
echoes in particle accelerators \citep{Stupakov1992,Spentzouris1996,Stupakov2013,Sen2018},
echoes in free-electron lasers \citep{Hemsing2014}, and echoes in
laser-kicked molecules \citep{Karras2015,Lin2016,Lin2020}. In addition,
echoes have been observed in single quantum systems, such as atoms
interacting with a quantized mode of electromagnetic radiation \citep{Morigi2002,Meunier2005},
and in single vibrationally excited molecules \citep{Qiang2020}.

In this paper, we study echoes in a single impulsively excited quantum
Kerr-nonlinear oscillator. In case of negligible damping and without
external drive, the system is modeled by the Hamiltonian \citep{Drummond1980}
\begin{equation}
\hat{H}_{0}=\hbar\omega\hat{a}^{\dagger}\hat{a}+\hbar\chi(\hat{a}^{\dagger})^{2}\hat{a}^{2},\label{eq:H0}
\end{equation}
where $\hat{a}^{\dagger}$ and $\hat{a}$ are the canonical creation
and annihilation operators (satisfying the commutation relation $[\hat{a},\hat{a}^{\dagger}]=1$),
$\omega$ is the fundamental frequency of the oscillator, and $\chi$
is the anharmonicity parameter. The echoes studied here do not require
inhomogeneous broadening in an ensemble of many oscillators. On contrary,
they occur in individual quantum systems, like in \citep{Morigi2002,Meunier2005,Qiang2020},
and completely rely on their intrinsic unitary dynamics.

The exactly solvable model described by the Hamiltonian in Eq. (\ref{eq:H0})
\citep{Milburn1986,Stobinska2008,Oliva2019}, and its driven damped
extensions \citep{Drummond1980} have been extensively studied theoretically.
For the recent theoretical developments in this direction, see \citep{Roberts2020}
and the references therein. In modern experiments, specially designed
superconducting quantum circuits allow studying the dynamics of the
dissipationless system {[}see Eq. (\ref{eq:H0}){]}, including the
phenomena of state collapse and quantum revivals \citep{Kirchmair2013}.
Such circuits are important in the context of quantum computing, and
are used in developing various state preparation \citep{Puri2017}
and state protection \citep{Ofek2016} protocols. For recent reviews
of quantum computing applications see, e.g. the references in \citep{Ofek2016,Puri2017}.
The Hamiltonian in Eq. (\ref{eq:H0}) also describes quantum non-linear
mechanical and opto-mechanical mesoscopic oscillatory systems \citep{NonOsc2012,QOptomech2020}.

The paper is organized as follows. In Sec. \ref{sec:The-Model}, we
introduce the interaction term used to model the impulsive excitation,
and define the corresponding classical Kerr-nonlinear oscillator model.
In Sec. \ref{sec:Echo-Effect}, the classical and quantum echoes are
presented, and the mechanisms of their formation are discussed. In
Sec. \ref{sec:Echo-Effect-Damped}, we consider the effects of dissipation.
Finally, Sec. \ref{sec: VII, Conclusions} concludes the paper.

\section{The Model \label{sec:The-Model}}

\subsection{Quantum model}

We begin by considering a quantized Kerr-nonlinear oscillator ``kicked''
by a pulsed coherent field. The system including the external field
is modeled by the Hamiltonian $\hat{H}=\hat{H}_{0}+\hat{H}_{\mathrm{int}}$,
where the interaction term is given by \citep{Drummond1980}
\begin{equation}
\hat{H}_{\mathrm{int}}=E_{0}f(t)(e^{-i\omega_{L}t}\hat{a}^{\dagger}+e^{i\omega_{L}t}\hat{a}).\label{eq:Hint}
\end{equation}
Here, $\omega_{L}$ is the carrier frequency of the external field,
$E_{0}$ is its peak amplitude, and $f(t)$ defines the time dependence
of the amplitude. The interaction term $\hat{H}_{\mathrm{int}}$ is
written under the rotating wave approximation, assuming $|\omega_{L}-\omega|\ll\omega$
\citep{Berman2011book}.

For convenience, we apply the unitary transformation $\ket{\psi}=e^{-i\omega_{L}\hat{a}^{\dagger}\hat{a}t}\ket{\psi_{L}}$,
which removes the oscillating factors, $\exp(\pm i\omega_{L}t)$ from
the Hamiltonian. The details are summarized in Appendix \ref{sec:App-Unitary-transformation}.
After the unitary transformation, the Hamiltonian reads \citep{Drummond1980}
\begin{equation}
\hat{H}_{L}=\hbar\Delta\hat{a}^{\dagger}\hat{a}+\hbar\chi(\hat{a}^{\dagger})^{2}\hat{a}^{2}+E_{0}f(t)(\hat{a}^{\dagger}+\hat{a}),\label{eq:rotated-H}
\end{equation}
where $\Delta=\omega-\omega_{L}$ is the detuning. To simplify the
notation, the subindex $L$ is omitted in the rest of the paper.

We introduce dimensionless parameters $\tilde{\Delta}=\Delta/\chi$,
$\tilde{E}_{0}=E_{0}/(\hbar\chi)$, and time $\tilde{t}=t\chi$, such
that the Hamiltonian in Eq. (\ref{eq:rotated-H}) becomes
\begin{equation}
\hat{\mathcal{H}}=\tilde{\Delta}\hat{a}^{\dagger}\hat{a}+(\hat{a}^{\dagger})^{2}\hat{a}^{2}+\tilde{E}_{0}\tilde{f}(\tilde{t})(\hat{a}^{\dagger}+\hat{a}),\label{eq:H-dimensionless}
\end{equation}
where energy is measured in units of $\hbar\chi$.

\subsection{Corresponding classical model}

A dimensionless classical Hamiltonian, formally corresponding to the
quantum one in Eq. (\ref{eq:H-dimensionless}), can be defined in
terms of the complex variable $a$ and its complex conjugate $a^{*}$,
$\mathcal{H}=\tilde{\Delta}a^{*}a+(a^{*})^{2}a^{2}+\tilde{E}_{0}\tilde{f}(\tilde{t})(a^{*}+a)$.
For mechanical systems, the variable $a$ is expressed in terms of
dimensionless classical coordinate $\tilde{q}$ and momentum $\tilde{p}$,
$a=(\tilde{q}+i\tilde{p})/\sqrt{2}$, where $\tilde{q}=q/\sqrt{\hbar/\omega m}$,
$\tilde{p}=p\sqrt{\hbar\omega m}$, $m$ is the mass of the oscillator,
and $\omega$ is the fundamental frequency of the oscillator. The
appearance of $\hbar$ in these definitions is purely to allow a convenient
comparison with the quantum case \citep{Milburn1986}. In terms of
$\tilde{q}$ and $\tilde{p}$, the classical Hamiltonian reads
\begin{equation}
\mathcal{H}=\frac{\tilde{\Delta}}{2}(\tilde{q}^{2}+\tilde{p}^{2})+\frac{1}{4}(\tilde{q}^{2}+\tilde{p}^{2})^{2}+\sqrt{2}\tilde{E}_{0}\tilde{f}(\tilde{t})\tilde{q}.\label{eq:classical-Hamiltonian}
\end{equation}
In the rest of this paper, we use the scalings defined in this section,
and omit the tildes to simplify the notation.

\section{Echo Effect: Negligible Damping \label{sec:Echo-Effect}}

In this section, we study echoes in an impulsively excited Kerr-nonlinear
oscillator. We begin from considering the free evolution of an oscillator
initially prepared in a coherent state. Then, we consider the case
when a pulsed excitation is applied after a delay $\tau$. The results
of direct numerical simulation are qualitatively explained and compared
with approximate analytical results.

\subsection{Evolution of a free oscillator \label{subsec:Free-evolution}}

\begin{figure}[H]
\begin{centering}
\includegraphics[width=8.4cm]{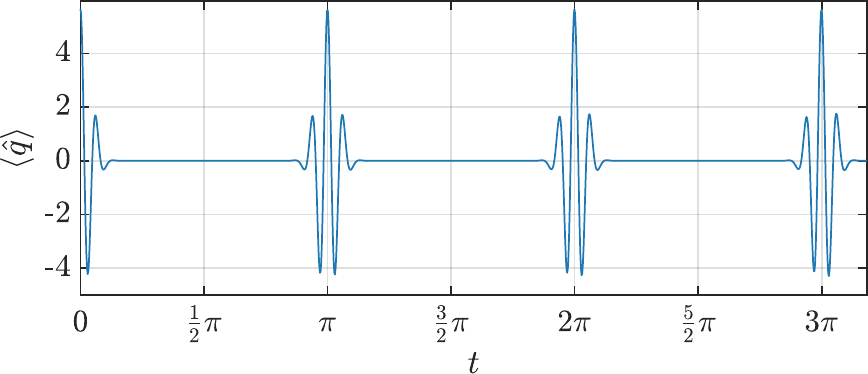}
\par\end{centering}
\caption{The function $\braket{\hat{q}}(t)$ {[}see Eq. (\ref{eq:expectation-q}){]}
with $\Delta=0$ and $\alpha_{0}=4$. \label{fig:free}}
\end{figure}

Initially, the oscillator is in a coherent state $\ket{\alpha_{0}}$.
In terms of number states $\{\ket{n}\}$ (defined by $\hat{a}^{\dagger}\hat{a}\ket{n}=n\ket{n}$)
it is given by
\begin{equation}
\ket{\psi(t=0)}=\ket{\alpha_{0}}=e^{-|\alpha_{0}|^{2}/2}\sum_{n=0}^{\infty}\frac{\alpha_{0}^{n}}{\sqrt{n!}}\ket{n},\label{eq:initial-state}
\end{equation}
where $\alpha_{0}$ is a complex number defining the coherent state.
The wave function of the oscillator at time $t$ reads
\begin{equation}
\ket{\psi(t)}=e^{-|\alpha_{0}|^{2}/2}\sum_{n=0}^{\infty}\frac{\alpha_{0}^{n}}{\sqrt{n!}}e^{-iE_{n}t}\ket{n},\label{eq:wave-packet}
\end{equation}
where $E_{n}=(\Delta-1)n+n^{2}$ is the energy of the state $\ket{n}$
{[}see Eq. (\ref{eq:H-dimensionless}), with $E_{0}=0${]}. To follow
the wave packet dynamics, we use the expectation value of the operator
$\hat{q}=(\hat{a}^{\dagger}+\hat{a})/\sqrt{2}$, which represents
position in the case of mechanical systems \citep{CTBook2020}, or
one of the field quadratures in the case of a quantized cavity mode
\citep{ScullyBook,SchleichBook}. It can be shown (see Appendix \ref{sec:App-Free-propagation})
that for the state in Eq. (\ref{eq:wave-packet}) the expectation
value $\braket{\hat{q}}(t)$ is given by
\begin{equation}
\braket{\hat{q}}(t)=\frac{e^{-|\alpha_{0}|^{2}}}{\sqrt{2}}\left[\alpha_{0}\exp(|\alpha_{0}|^{2}e^{-2it}-i\Delta\cdot t)+\mathrm{c.c}\right],\label{eq:expectation-q}
\end{equation}
where c.c. stands for ``complex conjugate''. Figure \ref{fig:free}
shows an example curve for $\braket{\hat{q}}(t)$ obtained using Eq.
(\ref{eq:expectation-q}) with $\Delta=0$, and $\alpha_{0}=4$. As
can be seen from the figure {[}and from Eq. (\ref{eq:expectation-q}){]},
the signal is $\pi$ periodic due to the quantum revivals \citep{Parker1986,Averbukh1989,Robinett2004}
the wave packet experiences at $t=kT_{rev}=k\pi,\,k=1,2,\dots$

Expanding the inner exponents, $\exp(\pm2it)$ in Eq. (\ref{eq:expectation-q})
up to the second order results in an approximate expression describing
the behavior of $\braket{\hat{q}}(t)$ in the vicinity of quantum
revivals ($t\approx k\pi,\,k=1,2,\dots$)
\begin{equation}
\braket{\hat{q}}(t)\approx\alpha_{0}e^{-2|\alpha_{0}|^{2}t^{2}}\cos(2t|\alpha_{0}|^{2}-\Delta\cdot t).\label{eq:free-short-1}
\end{equation}

The function in Eq. (\ref{eq:free-short-1}) oscillates at frequency
$2|\alpha_{0}|^{2}$ and decays to zero on a time-scale of $t_{c}=1/(2|\alpha_{0}|)$,
i.e. $\braket{\hat{q}}\propto\exp[-t^{2}/(2t_{c}^{2})]$, which is
the duration of the so-called ``wave packet collapse''. The reason
behind the collapse is the non-equidistant energy spectrum of the
Kerr-nonlinear oscillator. In other words, the frequencies of the
$\ket{n}$ states (forming the initial wave packet) are not integer
multiples of the fundamental frequency $\omega$, and, as a result,
the states quickly step out of phase.

\subsection{Evolution of an impulsively excited oscillator}

\begin{figure}
\begin{centering}
\includegraphics[width=0.95\columnwidth]{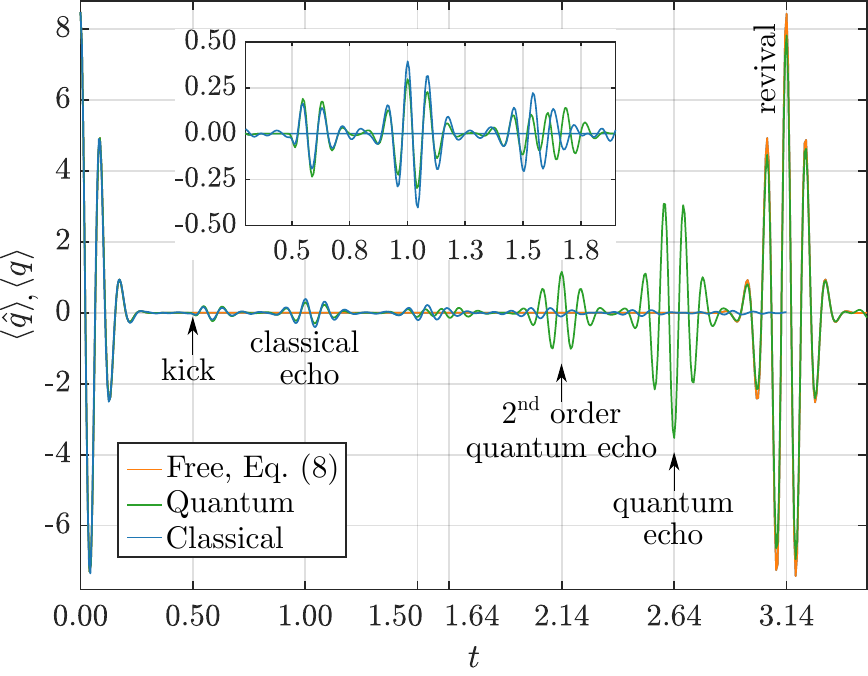}
\par\end{centering}
\caption{Quantum and classical echoes in kicked Kerr-nonlinear oscillator.
$\braket{\hat{q}}(t)$ is the quantum expectation value, while $\braket{q}(t)$
is the average position of $N=5\times10^{4}$ classical oscillators.
The detuning parameter is $\Delta=0.01$. The initial coherent state
is defined by $\alpha_{0}=6$. The kick is applied at $t=\tau=0.5$.
An immediate response to the excitation can be seen. The classical
echo emerges at $t=2\tau=1.0$. The quantum echo of the first order
appears at $t=T_{rev}-\tau=2.64$, while the second order quantum
echo is centered at $t=T_{rev}-2\tau=2.14$ ($T_{rev}=\pi$). Excitation
is Gaussian in time, $E_{0}f(t)=E_{0}\exp[-(t-\tau)^{2}/\sigma^{2}]$,
with $E_{0}=3.0$ and $\sigma=0.01$. The inset shows a magnified
portion of the plot. \label{fig:echo-classical}}
\end{figure}

Next, we consider results of an impulsive excitation (a ``kick'')
applied after a delay $\text{\ensuremath{\tau}}$ (counted from the
beginning of the evolution) to the oscillator being initially in a
coherent state $\ket{\alpha_{0}}$. Figure \ref{fig:echo-classical}
shows the expectation value $\braket{\hat{q}}(t)=\braket{\hat{a}^{\dagger}+\hat{a}}/\sqrt{2}$
calculated numerically in two different ways: (i) by solving the time-dependent
Shr\"odinger equation with Hamiltonian in Eq. (\ref{eq:H-dimensionless}),
and (ii) by simulating the behavior of a classical ensemble corresponding
to $\ket{\alpha_{0}}$. The curve describing the free propagation
{[}see Eq. (\ref{eq:expectation-q}){]} is added for comparison.

The behavior of the classical ensemble is studied with the help of
the Monte Carlo approach. Hamilton's equations of motion, derived
from the classical Hamiltonian in Eq. (\ref{eq:classical-Hamiltonian}),
are solved numerically for an ensemble of $N\gg1$ oscillators. The
initial position $q_{0}$ and momentum $p_{0}$ of the oscillators
are distributed according to
\begin{equation}
P(q_{0},p_{0})\propto\exp\left[-\frac{(q_{0}-\sqrt{2}\alpha_{0})^{2}}{2\sigma_{q}^{2}}-\frac{p_{0}^{2}}{2\sigma_{p}^{2}}\right],\label{eq:classical-initial-state}
\end{equation}
where $\sigma_{q}=\sigma_{p}=1/\sqrt{2}$. This initial classical
distribution corresponds to a coherent state $\ket{\alpha_{0}}$.
The classical observable $\braket{q}(t)$ is the average position
of the oscillators.

As seen in Fig. \ref{fig:echo-classical}, the quantum and classical
results are in good agreement during the initial stages of evolution.
The kick is applied at $t=\tau=0.5$, and both quantum and classical
simulations predict the expected immediate response to the applied
excitation. Later on, at twice the kick delay ($t=2\tau=1.0$), coherent
oscillations appear again without any additional kicks. On the long
time scale, several additional pulsed responses having remarkably
large amplitudes emerge at $t=T_{rev}-\tau=2.64$, and $t=T_{rev}-2\tau=2.14$
($T_{rev}=\pi$).

The described pulsed responses are similar to echo signals known in
many other physical systems. The characteristic property of echoes
is their timing---echoes appear at multiples of the kick delay, i.e.
$t=\pm l\tau\;\mathrm{mod}(T_{rev})$, where $l=1,2,\dots$ Here,
we refer to echoes at $t=+l\tau\;\mathrm{mod}(T_{rev})$ as ``classical
echoes of order $l$'', because they emerge in the quasi-classical
limit (see Fig. \ref{fig:echo-classical}, at $t=2\tau=1.0$) and
the mechanism behind their formation is classical. In contrast, echoes
at $t=-l\tau\;\mathrm{mod}(T_{rev})$ are of purely quantum origin.
They emerge due to the quantum revivals phenomenon, and we refer to
them as ``quantum echoes of order $l$'' (see Fig. \ref{fig:echo-classical},
at $t=T_{rev}-\tau=2.64$ and $t=T_{rev}-2\tau=2.14$). Quantum echoes
having similar timing were studied theoretically in ensembles of anharmonically
confined atoms \citep{Herrera2012}, and observed experimentally in
a gas of laser-kicked linear molecules \citep{Lin2016,Lin2017,Lin2020}.
Similar effects, but unrelated to the revivals phenomenon, were also
studied in ensembles of nonlinear systems with equidistant spectrum
\citep{Dubetskii1985,Dubetskii1986}.

Before proceeding further, it is important to emphasize the conceptual
difference between the echoes observed in inhomogeneous ensembles
of many particles and echoes in single particle systems \citep{Morigi2002,Meunier2005,Qiang2020}.
In the latter case, including the echo in a single mode of electromagnetic
field described by Hamiltonian in Eq. (\ref{eq:H-dimensionless}),
the effect does not require inhomogeneous broadening of the particle
properties, but relies solely on the quantum nature of the dynamics.

\subsection{Quantum echo - mechanism of formation \label{subsec:Mechanism-quantum-echo}}

To reveal the mechanism of echo formation, we begin by considering
the limit of weak impulsive excitations. In this limit, the temporal
extent of $f(t)$ {[}the function defining the time dependence of
the kick, see Eq. (\ref{eq:H-dimensionless}){]} is much shorter than
unity (in units of $\chi^{-1}$).

In the impulsive approximation, during the kick, the Hamiltonian in
Eq. (\ref{eq:H-dimensionless}) can be approximated by $\hat{\mathcal{H}}\approx E_{0}f(t)(\hat{a}^{\dagger}+\hat{a})$.
This allows to model the effect of the kick as $\ket{\psi_{+}}\approx\exp(\beta\hat{a}^{\dagger}-\beta^{*}\hat{a})\ket{\psi_{-}}$,
where $\ket{\psi_{\pm}}$ are the wave functions before/after the
kick, and $\beta=i\lambda=-iE_{0}\int_{-\infty}^{\infty}f(t)\,dt$.
The action of the kick is equivalent to application of the displacement
operator $\hat{D}(\beta)=\exp(\beta\hat{a}^{\dagger}-\beta^{*}\hat{a})$,
which shifts coherent states, $\hat{D}(\beta)\ket{\alpha}=\exp[(\alpha^{*}\beta-\alpha\beta^{*})/2]\ket{\alpha+\beta}$.

The kick is applied at $t=\tau$, and at that moment the state of
the oscillator reads
\begin{equation}
\ket{\psi_{-}}=e^{-|\alpha_{0}|^{2}/2}\sum_{n=0}^{\infty}\frac{\alpha_{0}^{n}}{\sqrt{n!}}e^{-iE_{n}\tau}\ket{n}.\label{eq:before-kick}
\end{equation}
Using the completeness property of coherent states, we expand each
number state $\ket{n}$ in terms of coherent states
\begin{equation}
\ket{n}=\frac{1}{\pi}\int_{\mathbb{C}}e^{-|\alpha|^{2}/2}\frac{(\alpha^{*})^{n}}{\sqrt{n!}}\ket{\alpha}\,d^{2}\alpha.\label{eq:number-state-expantion}
\end{equation}
Immediately after the kick, the state is given by
\begin{align}
\ket{\psi_{+}}\approx D(\beta)\ket{\psi_{-}} & =\frac{e^{-|\alpha_{0}|^{2}/2}}{\pi}\times\nonumber \\
\sum_{n=0}^{\infty}\frac{\alpha_{0}^{n}}{n!}e^{-iE_{n}\tau} & \int_{\mathbb{C}}e^{-|\alpha|^{2}/2}(\alpha^{*})^{n}\hat{D}(\beta)\ket{\alpha}\,d^{2}\alpha.\label{eq:after-kick}
\end{align}
Next, we expand each coherent state $\ket{\alpha+\beta}$ in terms
of number states $\{\ket{m}\}$, and substitute $\beta=i\lambda$
\begin{align}
\ket{\psi_{+}}=e^{-|\alpha_{0}|^{2}/2}\sum_{m,n=0}^{\infty}\frac{\alpha_{0}^{n}}{n!}\frac{1}{\sqrt{m!}}e^{-iE_{n}\tau}\ket{m} & \times\nonumber \\
\frac{1}{\pi}\int_{\mathbb{C}}e^{-|\alpha|^{2}}(\alpha^{*})^{n}e^{i\alpha\lambda-\lambda^{2}}(\alpha+i\lambda)^{m}\,d^{2}\alpha.\label{eq:wf-after-kick}
\end{align}
Up to the first order in $\lambda$ (i.e. assuming $|\lambda|=E_{0}\int_{-\infty}^{\infty}f(t)\,dt\ll1$)
the complex integral in Eq. (\ref{eq:wf-after-kick}) becomes
\begin{align}
\frac{1}{\pi}\int_{\mathbb{C}}e^{-|\alpha|^{2}}(\alpha^{*})^{n}[\alpha^{m}+i\lambda m\alpha^{m-1}+i\lambda\alpha^{m+1}]\,d^{2}\alpha.\label{eq:integral-lambda}
\end{align}
The three resulting integrals can be evaluated by substituting $\alpha=re^{i\varphi}$,
changing variables $t=r^{2}$, and using the definition of the Gamma
function $\Gamma(n)=\int_{0}^{\infty}e^{-t}t^{2n}\,dt=(n-1)!$.

The state at time $T$ (counted from the moment of the kick), is given
by
\begin{equation}
\ket{\psi(T)}=\ket{\psi_{f}(T)}+\ket{\psi_{-1}(T)}+\ket{\psi_{1}(T)},\label{eq:wave-function-components-1}
\end{equation}
where
\begin{equation}
\begin{aligned}\ket{\psi_{f}(T)} & =A\sum_{n=0}^{\infty}\frac{\alpha_{0}^{n}}{\sqrt{n!}}e^{-iE_{n}(\tau+T)}\ket{n},\\
\ket{\psi_{-1}(T)} & =A\frac{i\lambda}{\alpha_{0}}\sum_{n=0}^{\infty}\frac{n\alpha_{0}^{n}}{\sqrt{n!}}e^{-i(E_{n-1}\tau+E_{n}T)}\ket{n},\\
\ket{\psi_{1}(T)} & =Ai\lambda\alpha_{0}\sum_{n=0}^{\infty}\frac{\alpha_{0}^{n}}{\sqrt{n!}}e^{-i(E_{n+1}\tau+E_{n}T)}\ket{n},
\end{aligned}
\label{eq:wave-function-components-2}
\end{equation}
and $A=\exp(-|\alpha_{0}|^{2}/2)$. After the weak kick, the state
of the oscillator is composed of three wave packets. The first one,
$\ket{\psi_{f}}$ is identical to the freely evolving initial coherent
state $\ket{\alpha_{0}}$ {[}see Eq. (\ref{eq:wave-packet}){]}. Using
the expression for energy $E_{n}=(\Delta-1)n+n^{2}$, the newly created
wave packets read
\begin{equation}
\begin{aligned}\ket{\psi_{-1}(T)} & \propto\sum_{n=0}^{\infty}\frac{n[\alpha_{0}e^{2i\tau}]^{n}}{\sqrt{n!}}e^{-iE_{n}(\tau+T)}\ket{n},\\
\ket{\psi_{1}(T)} & \propto\sum_{n=0}^{\infty}\frac{[\alpha_{0}e^{-2i\tau}]^{n}}{\sqrt{n!}}e^{-iE_{n}(\tau+T)}\ket{n}.
\end{aligned}
\label{eq:new-packets}
\end{equation}

This shows that $\ket{\psi_{1}}$ has the form of a coherent state
$\ket{\alpha_{0}\exp(-2i\tau)}$ propagating freely for the time $\tau+T$.

\begin{figure}
\begin{centering}
\includegraphics[width=8.4cm]{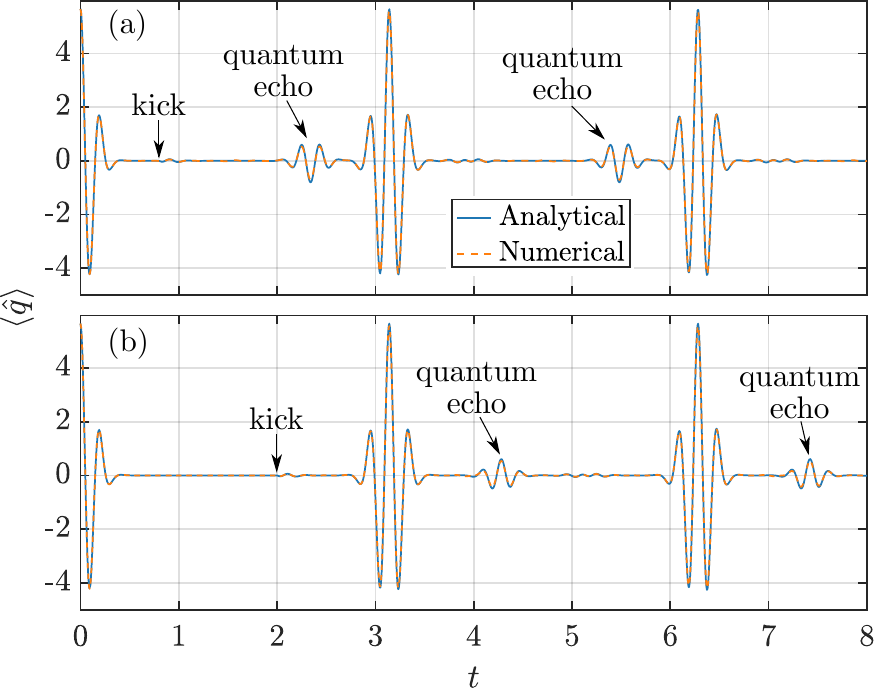}
\par\end{centering}
\caption{Comparison of the expectation value $\braket{\hat{q}}(t)$ obtained
using Eq. (\ref{eq:q-after-kick-lambda}) (solid blue), and numerically
(dashed orange). Detuning parameter is $\Delta=0.01$. The initial
coherent state is defined by $\alpha_{0}=4$. The kick is applied
at (a) $t=\tau=0.8$, and (b) $t=\tau=2.0$, while the first echo
response emerges at $t=T_{rev}-\tau=2.34$ and $t=2T_{rev}-\tau=4.28$,
respectively. Excitation is Gaussian in time, $E_{0}f(t)=E_{0}\exp[-(t-\tau)^{2}/\sigma^{2}]$,
with $E_{0}=0.50$, and $\sigma=0.02$. \label{fig:echo-analytical-1}}
\end{figure}

\begin{figure}
\begin{centering}
\includegraphics[width=0.95\columnwidth]{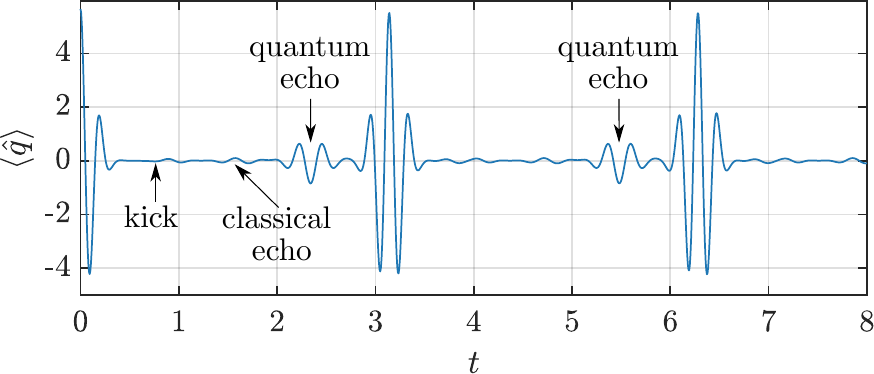}
\par\end{centering}
\caption{Numerically calculated quantum expectation value $\braket{\hat{q}}(t)$.
The detuning parameter is $\Delta=0.01$. The initial coherent state
is defined by $\alpha_{0}=4$. The kick is applied at $t=\tau=0.8$,
and the immediate response to the excitation can be seen. Low-amplitude
classical echo is visible at $t=2\tau=1.6$, while the quantum echo
appears at $t=T_{rev}-\tau=2.14$. Excitation is Gaussian in time,
$E_{0}f(t)=E_{0}\exp[-(t-\tau)^{2}/\sigma^{2}]$, with $E_{0}=1.0$,
and $\sigma=0.1$ \label{fig:echo-classical-1}}
\end{figure}

\begin{figure*}
\begin{centering}
\includegraphics[width=0.95\textwidth]{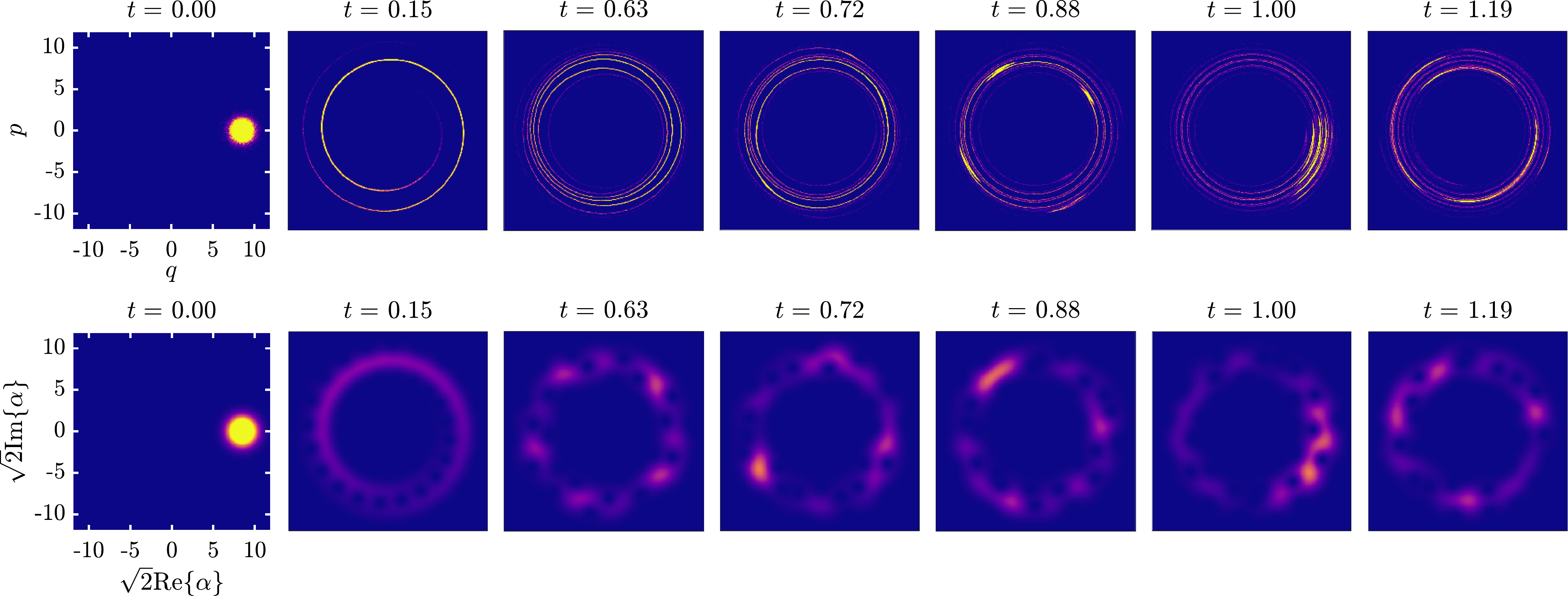}
\par\end{centering}
\caption{Top row - snapshots of the classical phase space distribution. Bottom
row - corresponding Husimi Q-distribution. The detuning parameter
is $\Delta=0.01$. The initial coherent state is defined by $\alpha_{0}=6$.
The kick is applied at $t=\tau=0.50$. The snapshot at $t=2\tau=1.0$,
shows how the bunches in the classical phase space distribution synchronize,
producing the classical echo. Similar synchronization phenomenon is
seen in the Q-distribution at the same moment. Excitation is Gaussian
in time, $E_{0}f(t)=E_{0}\exp[-(t-\tau)^{2}/\sigma^{2}]$, with $E_{0}=15$,
and $\sigma=0.01$. \label{fig:phase-space}}
\end{figure*}

We proceed with the analysis of the expectation value of the operator
$\hat{q}=(\hat{a}^{\dagger}+\hat{a})/\sqrt{2}$. It can be shown (see
Appendix \ref{sec:App-Second-order-lambda}) that up to the terms
first order in $\lambda$, $\braket{\hat{q}}(t)$ is given by
\begin{equation}
\begin{aligned}\braket{\hat{q}}(t) & =c_{f}(t)e^{z(t)}\\
 & +\lambda c_{-1}(t)e^{z(t-\tau)}+\lambda c_{1}(t)e^{z(t+\tau)}\\
 & +\mathrm{c.c.},
\end{aligned}
\label{eq:q-after-kick-lambda}
\end{equation}
where ``c.c.'' stands for complex conjugate, $t$ is time counted
from the beginning of the evolution, $z(t)=|\alpha_{0}|^{2}e^{-2it}$,
and $c_{f}(t)=\alpha_{0}\exp(-|\alpha_{0}|^{2}-i\Delta\cdot t)/\sqrt{2}$.
The first term in Eq. (\ref{eq:q-after-kick-lambda}) is identical
to $\braket{\hat{q}}(t)$ of the freely evolving initial coherent
state $\ket{\alpha_{0}}$ {[}see Eq. (\ref{eq:expectation-q}){]},
while the other two stem from the quantum interference of $\ket{\psi_{f}}$
and $\ket{\psi_{\pm1}}$ {[}see Eq. (\ref{eq:wave-function-components-2}){]}.
The time-dependent coefficients $c_{\pm1}$ are defined in Appendix
\ref{sec:App-Second-order-lambda} {[}see Eq. (\ref{eq:App-c-coefficients}){]}.
The overall behavior of the two terms proportional to $\lambda$ is
determined by the functions $\exp[z(t\pm\tau)]$, which have a characteristic
width of $t_{c}=1/(2|\alpha_{0}|)$ {[}see Eqs. (\ref{eq:expectation-q})
and (\ref{eq:free-short-1}){]}.

The function $\exp[z(t-\tau)]$ is centered at $t=\tau$ (the moment
of the kick) and describes the expected impulsive response to the
kick, while $\exp[z(t+\tau)]$ is centered at $t=-\tau$. Both functions
are $\pi$ periodic, and therefore $\exp[z(t+\tau)]$ describes localized
oscillations emerging periodically \emph{before each revival,} at
$t=kT_{rev}-\tau$, $k=1,2,\dots$ This delayed response is the previously
discussed quantum echo of the first order (see Fig. \ref{fig:echo-classical}
at $t=T_{rev}-\tau=2.64$).

Figure \ref{fig:echo-analytical-1} shows two examples in which the
kick is applied at $\tau=0.8$ {[}panel (a){]}, and $\tau=2.0$ {[}panel
(b){]}. In Fig. \ref{fig:echo-analytical-1}(a), the quantum echo
appears for the first time before the first revival, at $t=T_{rev}-\tau=\pi-\tau=2.34$.
In Fig. \ref{fig:echo-analytical-1}(b), the timing of the kick is
such that the echo appears only before the second revival, at $t=2T_{rev}-\tau=2\pi-\tau=4.28$.
For the chosen kick parameters, the analytical result {[}see Eq. (\ref{eq:q-after-kick-lambda}){]}
and the result of direct numerical solution of the time-dependent
Shr\"odinger equation are in good agreement.

So far, we have demonstrated that the quantum echo at $t=T_{rev}-\tau$
is a first order effect in $\lambda$ (the excitation strength parameter).
In the next section, we show that the classical echo at $t=2\tau$
{[}see Fig. (\ref{fig:echo-classical}){]} emerges only in the second
order of $\lambda$, which explains the dominating amplitude of the
quantum echo.

As an additional numerical example, we consider the case of a kick
having a slightly higher amplitude ($E_{0}=1.0$) and longer duration
($\sigma=0.1$), which can not be treated within the impulsive approximation
used to derive Eq. (\ref{eq:q-after-kick-lambda}). Figure \ref{fig:echo-classical-1}
shows that the first order quantum echo still appears before the quantum
revival, and also a weak classical echo is visible at $t=2\tau$.

\subsection{Classical echo - mechanism of formation \label{subsec:Mechanism-classical-echo}}

In this section, we consider the effects up to the second order in
the kick strength $\lambda$, and discuss the emergence of the classical
echo. It can be shown (see Appendix \ref{sec:App-Second-order-lambda})
that up to the second order in $\lambda$, the expectation value of
the operator $\hat{q}=(\hat{a}^{\dagger}+\hat{a})/\sqrt{2}$ is given
by
\begin{equation}
\begin{aligned}\braket{\hat{q}}(t) & =[c_{f}(t)+\lambda^{2}c_{0}(t)]e^{z(t)}\\
 & +\lambda\sum_{j=\pm1}c_{j}(t)e^{z(t+j\tau)}\\
 & +\lambda^{2}\sum_{j=\pm2}c_{j}(t)e^{z(t+j\tau)}+\mathrm{c.c.},
\end{aligned}
\label{eq:q-after-kick-lambda2}
\end{equation}
where ``c.c.'' stands for complex conjugate, $\lambda=-E_{0}\int_{-\infty}^{\infty}f(t)\,dt$,
$z(t)=|\alpha_{0}|^{2}e^{-2it}$, and $c_{f}=\alpha_{0}\exp(-|\alpha_{0}|^{2}-i\Delta\cdot t)/\sqrt{2}$.
The terms proportional to $\lambda^{2}$ introduce two new time points
when the signal exhibits pulsed responses (echoes). The term $\exp[z(t+2\tau)]$
corresponds to second order quantum echo appearing before each revival\emph{,}
at $t=kT_{rev}-2\tau$, $k=1,2,\dots$ The term $\exp[z(t-2\tau)]$
corresponds to a response appearing after an additional delay $\tau$
after the kick, at $t=2\tau$. This is the classical echo of the first
order. The time-dependent coefficients $c_{\pm1,2}$ are defined in
Appendix \ref{sec:App-Second-order-lambda} {[}see Eq. (\ref{eq:App-c-coefficients}){]}.

We can gain a physical insight into the mechanism of the classical
echo formation by considering the classical phase space dynamics.
Figure \ref{fig:phase-space} shows a series of snapshots of the classical
phase space distribution at several times before and after the kick
applied at $t=\tau=0.5$. For comparison, the lower row shows the
corresponding quantum Husimi Q-distribution, defined as \citep{SchleichBook}
\[
Q(q,p,t)=Q(\sqrt{2}\mathrm{Re}\{\alpha\},\sqrt{2}\mathrm{Im}\{\alpha\},t)=\frac{1}{\pi}\braket{\alpha|\hat{\rho}|\alpha},
\]
where $\hat{\rho}$ is the density matrix of the oscillator (for a
pure quantum state $\hat{\rho}=\ket{\psi}\bra{\psi}$). In an anharmonic
oscillator, the frequency (period) of oscillations is energy-dependent,
or in other words, the period of oscillations depends on the radial
distance from the phase space origin. As a result, with time, the
initial smooth distribution at $t=0$ evolves into a spiral-like structure
(seen at $t=0.15$). This filamentation of the phase space results
in an increasing number of spiral turns, which become thinner in order
to conserve the initial phase space volume (Liouville's theorem).
Filamentation of the phase space is a known nonlinear phenomenon,
which emerges in various physical systems, e.g. in the dynamics of
stellar systems \citep{Lynden-Bell1967}, and in accelerator physics
\citep{Guignard1988,Stupakov1992,Stupakov2013}.

The filamented phase space serves as a basis for the echo formation
induced by a kick applied at the moment of well developed filamentation
(at $t=\tau=0.5$). The kick suddenly shifts the phase space distribution
along the momentum axis, leading, with time, to the appearance of
density bunches on the spiral. The panels corresponding to $t=0.15,0.63,0.72$
show the creation of the bunches. As the time goes on, the bunches
evolve into sharp ``tips''. The tips are located on different turns
of the spiral, therefore they rotate with different frequencies and
step out of phase with time. However, at twice the kick delay (at
$t=2\tau=1.0$) the bunches/tips synchronize. This synchronization
is manifested as the classical echo effect \citep{Stupakov1993,Stupakov2013,Lin2016,Lin2020}.
The similarity between the behavior of the classical phase space distribution
and the Q-distribution is evident in Fig. \ref{fig:phase-space}.
In principle, the synchronization recurs periodically with a period
of $\tau$. The so-called higher order classical echoes \citep{Stupakov2013,Karras2015,Lin2016}
(partially visible in Fig. \ref{fig:echo-classical} at $t=3\tau=1.5$)
can be observed at higher multiples of the kick delay $t=3\tau,4\tau,\dots$

\begin{figure}
\begin{centering}
\includegraphics[width=0.95\columnwidth]{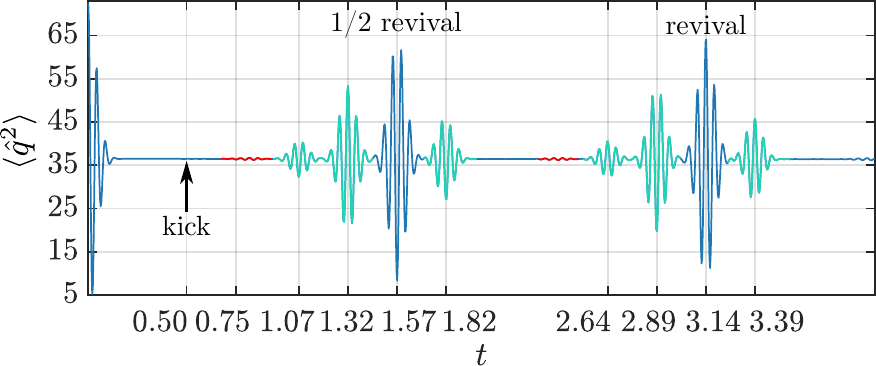}
\par\end{centering}
\caption{Fractional echoes in the expectation value $\braket{\hat{q}^{2}}$.
The kick is applied at $t=\tau=0.5$. All the simulation parameters
are identical to Fig. \ref{fig:echo-classical}. Regular and fractional
quantum echoes, marked in green, are visible at $t=T_{rev}/2\pm\tau/2,T_{rev}/2-\tau,T_{rev}\pm\tau/2,T_{rev}-\tau$.
Low-amplitude classical echo, marked in red, is visible at $t\approx3\tau/2$.
\label{fig:fractional-echoes}}
\end{figure}

As a final remark for this section, we would like to point out the
existence of the so-called ``fractional echoes'' in our system.
These echoes are visible when higher moments of the field distribution
are considered, e.g. $\braket{\hat{q}^{n}}\,n\geq2$ (see e.g. \citep{Karras2016,Lin2016,Wang2019}
and references therein). These echoes appear at rational fractions
of the kick delay $k\tau/l$ (where $k$ and $l$ are mutually prime
numbers) after the kick, and before/after revivals of various orders.
Figure \ref{fig:fractional-echoes} shows the quantum expectation
value of the operator $\hat{q}^{2}=(\hat{a}^{\dagger}+\hat{a})^{2}/2$
demonstrating the regular and fraction quantum echoes at $t=T_{rev}/2\pm\tau/2,T_{rev}/2-\tau,T_{rev}\pm\tau/2,T_{rev}-\tau$,
as well as low-amplitude classical fractional echo at $t\approx3\tau/2$.

\section{Echo Effect:dissipation effects \label{sec:Echo-Effect-Damped}}

In this section, we consider the role of damping effects caused by
interaction of the oscillator with a reservoir at finite temperature
$T$. The whole system (oscillator + reservoir) is described by the
time-dependent density matrix $\hat{\rho}(t)$. The oscillator is
described by the reduced density matrix $\hat{S}(t)=\mathrm{Tr}_{R}\hat{\rho}(t)$,
where $\mathrm{Tr}_{R}\cdot$ denotes partial trace over the reservoir's
degrees of freedom. We use the quantum optics model of a single mode
of Kerr resonator interacting with a reservoir. In this model, $\hat{S}(t)$
satisfies the following differential equation \citep{Louisell1973,Drummond1980}
\begin{align}
\frac{\partial\hat{S}}{\partial t} & =-i[\mathcal{\hat{H}},\hat{S}]+\gamma\bar{n}\left[[\hat{a},\hat{S}],\hat{a}^{\dagger}\right]\nonumber \\
 & +\frac{\gamma}{2}(2\hat{a}\hat{S}\hat{a}^{\dagger}-\hat{a}^{\dagger}\hat{a}\hat{S}-\hat{S}\hat{a}^{\dagger}\hat{a}),\label{eq:equation-S(t)}
\end{align}
where $\hat{\mathcal{H}}$ is given in Eq. (\ref{eq:H-dimensionless}),
$[\hat{A},\hat{B}]=\hat{A}\hat{B}-\hat{B}\hat{A}$, $\gamma$ is the
dimensionless damping constant, $\bar{n}=[\exp(\epsilon)-1]^{-1}$
{[}where $\epsilon=\hbar\omega/(k_{B}T)]$ is the mean number of bosonic
excitations in the reservoir's mode having frequency $\omega$ (the
fundamental frequency of the quantized mode of the field), and $k_{B}$
is the Boltzmann constant.

\begin{figure}
\begin{centering}
\includegraphics[width=0.95\columnwidth]{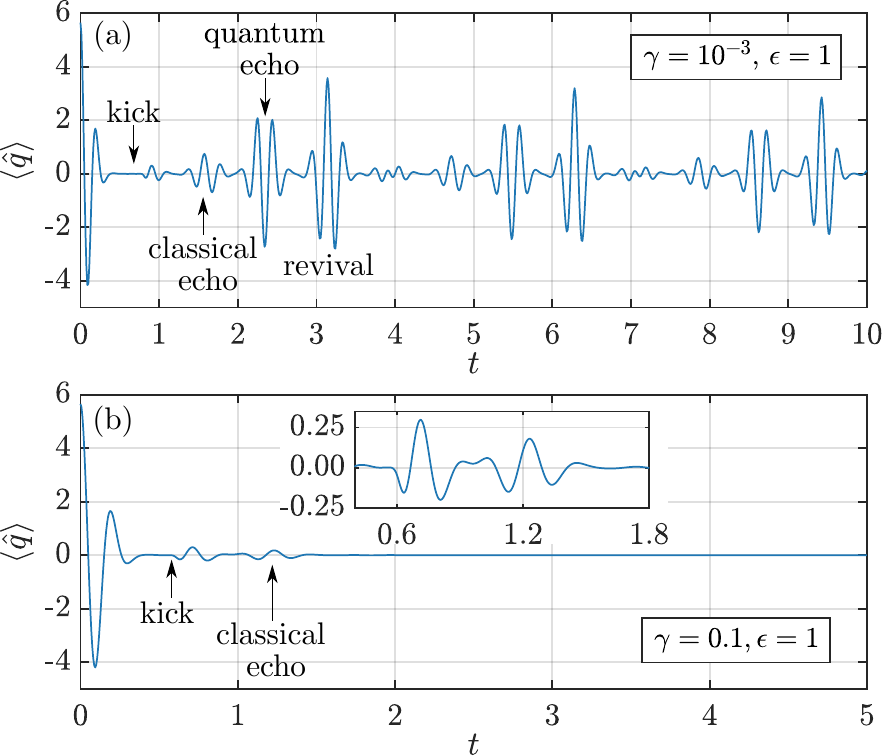}
\par\end{centering}
\caption{Numerically calculated quantum expectation value of the field quadrature
operator $\hat{q}=(\hat{a}^{\dagger}+\hat{a})/\sqrt{2}$ for the case
of initially coherent state defined by $\alpha_{0}=4$. The detuning
parameter is $\Delta=0.01$. (a) Weak damping ($\gamma=10^{-3}$),
the kick is applied at $t=\tau=0.8$. (b) Strong damping ($\gamma=0.1$),
the kick is applied at $t=\tau=0.6$. In both cases, the excitations
are Gaussian in time, $E_{0}f(t)=E_{0}\exp[-(t-\tau)^{2}/\sigma^{2}]$,
with $E_{0}=3.0$, and $\sigma=0.02$. \label{fig:damped-coherent}}
\end{figure}

Figure \ref{fig:damped-coherent} shows two examples of the damped
dynamics of $\braket{\hat{q}}$ for the case of a field being initially
in a coherent state $\ket{\alpha_{0}}$ {[}i.e. $\hat{S}(t=0)=\ket{\alpha_{0}}\bra{\alpha_{0}}${]}.
For weak damping, {[}$\gamma=10^{-3}$, see Fig. \ref{fig:damped-coherent}(a){]},
the amplitude of all oscillations (revivals and echoes) gradually
diminish with each revival cycle. In the case of stronger damping,
$[\gamma=0.1$, see Fig. \ref{fig:damped-coherent}(b){]}, even the
first revival at $t=T_{rev}=\pi$ is not visible. In contrast, the
classical echo appearing on the short time scale is clearly visible.
Since the quantum echo appears on the longer time scale (just before
the revival), its amplitude is negligible as compared to the amplitude
of the classical echo.

\begin{figure}
\begin{centering}
\includegraphics[width=0.95\columnwidth]{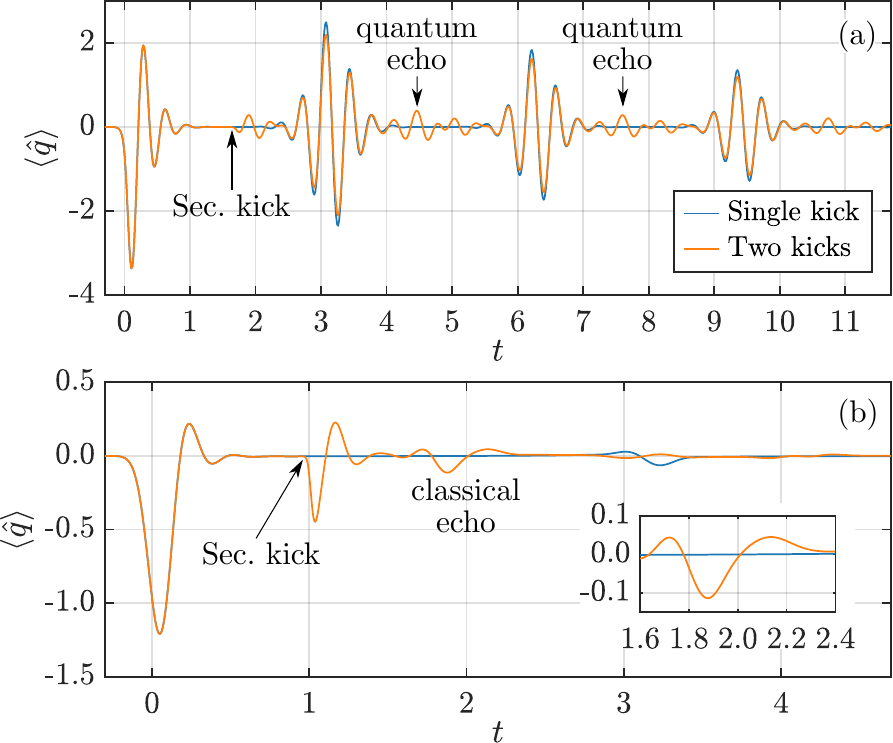}
\par\end{centering}
\caption{Numerically calculated quantum expectation value of the field quadrature
operator $\hat{q}=(\hat{a}^{\dagger}+\hat{a})/\sqrt{2}$ for the case
of initially thermal state and $\gamma=5\times10^{-3}$. The detuning
parameter is $\Delta=0.01$. The excitations are Gaussian, and the
time dependent amplitude is given by: $E_{01}\exp[-t^{2}/\sigma_{1}^{2}]+E_{02}\exp[-(t-\tau)^{2}/\sigma_{2}^{2}]$.
(a) Low temperature $\epsilon=1$, the second kick is applied at $t=\tau=1.7$,
with $E_{01}=20,\,\sigma_{1}=0.1$ and $E_{02}=1,\,\sigma_{2}=0.05$.
(b) High temperature $\epsilon=0.1$, the second kick is applied at
$t=\tau=1.0$, with $E_{01}=20,\,\sigma_{1}=0.1$ and $E_{02}=13,\,\sigma_{2}=0.02$.
\label{fig:damped-thermal}}
\end{figure}

As an additional example, we consider the case of a field being initially
in thermal equilibrium with the reservoir, i.e. the initial density
matrix of the quantized mode is given by
\begin{equation}
\hat{S}(t=0)=\sum_{n=0}^{\infty}P_{n}\ket{n}\bra{n}\quad P_{n}=\frac{1}{1+\bar{n}}\left(\frac{\bar{n}}{1+\bar{n}}\right)^{n}\label{eq:initial-thermal-state}
\end{equation}
where $\{P_{n}\}$ is the Bose-Einstein distribution. Figure \ref{fig:damped-thermal}
shows the damped dynamics of $\braket{\hat{q}}$, for both low and
high temperature reservoirs, and $\gamma=5\times10^{-3}$. Here, we
apply two delayed kicks to the system. The first one is stronger and
it initiates the dynamics, while the second delayed kick induces the
echoes.

In the case of low temperature {[}see Fig. \ref{fig:damped-thermal}(a){]},
the first kick at $t=0$ is followed by decaying oscillations of $\braket{\hat{q}}$
with each revival. The timing of the second kick is such that the
first quantum echo appears at $t=2T_{rev}-\tau$, similar to Fig.
\ref{fig:echo-analytical-1}(b). For high temperature {[}see Fig.
\ref{fig:damped-thermal}(b){]}, the oscillatory response to the first
kick decays faster as compared to the previous case, due to wider
distribution of initially populated states of the field. The second
kick is applied after the response to the first kick has been decayed,
and the classical echo arises at twice the kick delay, $t\approx2\tau$.
Since the contact with the hot reservoir leads to faster decoherence,
even the first quantum revival is not visible in this case.

\section{Conclusions \label{sec: VII, Conclusions}}

Two types of echoes, quantum and classical ones, are demonstrated
in the kicked Kerr-nonlinear oscillator. The classical echoes exist
on the short time-scale in the semi-classical limit, while the quantum
ones show themselves near the quantum revivals of various order. For
weak excitation, the amplitude of the quantum echoes is much higher
than the classical ones. The reason is that the quantum echo emerges
in the first order perturbation theory in the excitation strength,
while the classical echo first appears in the second order.

The echoes discussed in the present paper are somehow different from
echoes emerging in inhomogeneously broadened ensembles of many particles,
e.g. spin echoes \citep{Hahn1950PR,Hahn1953}. Here, they take place
in a single quantum object, and, in principle, can be observed in
a single quantum oscillator. In this case, the measurement should
be repeated many times starting from the same initial condition. The
interference echo structure developing after many measurements is
a time-domain analogue of spatial interference pattern appearing on
a screen as a result of sending single electrons one-by-one through
a double slit (the famous Feynman gedanken experiment, see \citep{Bach2013}
and references therein). Recently, a related phenomenon of echoes
in single vibrationally excited molecules was observed \citep{Qiang2020}.

The amplitudes of the various echoes are sensitive to the decoherence
processes. In the presence of strong damping, the quantum echoes and
revivals may be completely suppressed, because they emerge on the
long time-scale. In contrast, classical echoes appear on the short
time-scale and may still be measured. A similar situation was recently
encountered in experiments \citep{Rosenberg2018,Zhang2019,Ma2019,Hartmann2020}
studying collisional rotational relaxation in dense molecular gases.
Although the rotational revivals were essentially suppressed at high
pressure, the classical alignment echoes were readily observable and
successfully used to measure the relaxation rates.

\section*{Acknowledgments}

This work was partially supported by the Israel Science Foundation
(Grant No. 746/15). I.A. acknowledges support as the Patricia Elman
Bildner Professorial Chair. This research was made possible in part
by the historic generosity of the Harold Perlman Family.

\appendix

\section{Unitary transformation {[}Eq. (\ref{eq:rotated-H}){]} \label{sec:App-Unitary-transformation}}

The unitary transformation $\ket{\psi}=e^{-i\omega_{L}\hat{N}t}\ket{\psi_{L}}$,
where $\hat{N}=\hat{a}^{\dagger}\hat{a}$, allows removing the oscillating
factors, $\exp(\pm i\omega_{L}t)$ from the interaction term $\hat{H}_{\mathrm{int}}$
in Eq. (\ref{eq:Hint}). It can be shown that $\ket{\psi_{L}}$ satisfies
the Shr\"odinger equation with the Hamiltonian
\begin{equation}
\hat{H}_{L}=\hbar\Delta\hat{a}^{\dagger}\hat{a}+\hbar\chi(\hat{a}^{\dagger})^{2}\hat{a}^{2}+e^{i\omega_{L}\hat{N}t}\hat{H}_{\mathrm{int}}e^{-i\omega_{L}\hat{N}t},\label{eq:H_L-1}
\end{equation}
where $\Delta=\omega-\omega_{L}$ is the detuning. The time-dependent
operators $\hat{a}_{L}^{\dagger}(t)=\exp(i\omega_{L}\hat{N}t)\hat{a}^{\dagger}\exp(-i\omega_{L}\hat{N}t)$
and $\hat{a}_{L}(t)=\exp(i\omega_{L}\hat{N}t)\hat{a}\exp(-i\omega_{L}\hat{N}t)$
appearing in Eq. (\ref{eq:H_L-1}) satisfy the differential equations
\begin{equation}
\begin{aligned}\frac{\partial\hat{a}_{L}^{\dagger}(t)}{\partial t} & =i\omega_{L}e^{i\omega_{L}\hat{N}t}[\hat{N},\hat{a}^{\dagger}]e^{-i\omega_{L}\hat{N}t},\\
\frac{\partial\hat{a}_{L}(t)}{\partial t} & =i\omega_{L}e^{i\omega_{L}\hat{N}t}[\hat{N},\hat{a}]e^{-i\omega_{L}\hat{N}t},
\end{aligned}
\label{eq:App-adaga-diff-eq}
\end{equation}
having solutions $\hat{a}_{L}^{\dagger}(t)=\hat{a}^{\dagger}\exp(i\omega_{L}t)$
and $\hat{a}_{L}(t)=\hat{a}\exp(-i\omega_{L}t)$. Substituting these
into the Hamiltonian in Eq. (\ref{eq:H_L-1}), results in
\begin{equation}
\hat{H}_{L}=\hbar\Delta\hat{a}^{\dagger}\hat{a}+\hbar\chi(\hat{a}^{\dagger})^{2}\hat{a}^{2}+E_{0}f(t)(\hat{a}^{\dagger}+\hat{a}),\label{eq:App-rotated-H}
\end{equation}
which is the Hamiltonian in Eq. (\ref{eq:rotated-H}).

\section{Free propagation of a coherent state {[}Eq. (\ref{eq:expectation-q}){]}
\label{sec:App-Free-propagation}}

In this Appendix, we calculate the time-dependent quantum expectation
value of the operator $\hat{q}=(\hat{a}^{\dagger}+\hat{a})/\sqrt{2}$
for the free oscillator being initially in a coherent state $\ket{\alpha_{0}}$
{[}see Eq. (\ref{eq:initial-state}){]}. The wave function at time
$t$ reads
\begin{equation}
\ket{\psi(t)}=e^{-|\alpha_{0}|^{2}/2}\sum_{n=0}^{\infty}\frac{\alpha_{0}^{n}}{\sqrt{n!}}e^{-iE_{n}t}\ket{n},\label{eq:App-in-state-at-tau}
\end{equation}
where $E_{n}=(\Delta-1)n+n^{2}$. Applying the operator $\hat{a}$
to the wave function yields
\begin{equation}
\hat{a}\ket{\psi(t)}=e^{-|\alpha_{0}|^{2}/2}\sum_{n=1}^{\infty}\frac{\alpha_{0}^{n}}{\sqrt{n!}}e^{-iE_{n}t}\sqrt{n}\ket{n-1}.\label{eq:App-in-state-apply-a}
\end{equation}
For convenience, we change the index of summation $n$ to $k=n-1$,
such that
\begin{equation}
\hat{a}\ket{\psi(t)}=e^{-|\alpha_{0}|^{2}/2}\sum_{k=0}^{\infty}\frac{\alpha_{0}^{k+1}}{\sqrt{k!}}e^{-iE_{k+1}t}\ket{k}.\label{eq:App-in-state-index-change}
\end{equation}
The expectation value $\braket{\psi(t)|\hat{a}|\psi(t)}$ is given
by
\begin{align}
\alpha_{0}e^{-|\alpha_{0}|^{2}-i\Delta\cdot t}\sum_{n=0}^{\infty}\frac{\left(|\alpha_{0}|^{2}e^{-2it}\right)^{n}}{n!} & \ket{n}=\nonumber \\
\alpha_{0}e^{-|\alpha_{0}|^{2}}\exp(|\alpha_{0}|^{2}e^{-2it}-i\Delta\cdot t)\label{eq:App-free-expectation-a}
\end{align}
where we used $E_{n+1}-E_{n}=\Delta+2n$. The expectation value of
$\braket{\psi(t)|\hat{a}^{\dagger}|\psi(t)}$ is the complex conjugate
of Eq. (\ref{eq:App-free-expectation-a}). Finally, we have
\begin{align}
\braket{\hat{q}} & =\frac{1}{\sqrt{2}}\braket{\hat{a}^{\dagger}+\hat{a}}\nonumber \\
 & =\frac{e^{-|\alpha_{0}|^{2}}}{\sqrt{2}}\left[\alpha_{0}\exp(|\alpha_{0}|^{2}e^{-2it}-i\Delta\cdot t)+\mathrm{c.c}\right],\label{eq:App-expectation-of-q}
\end{align}
where c.c. stands for ``complex conjugate''.

\section{Derivation of Eq. (\ref{eq:q-after-kick-lambda2}) \label{sec:App-Second-order-lambda}}

In this Appendix, we derive the formula for $\braket{\hat{q}}(t)=\braket{\hat{a}^{\dagger}+\hat{a}}/\sqrt{2}$
up to the second order in $\lambda$ (the excitation strength parameter)
for an oscillator being initially in a coherent state $\ket{\alpha_{0}}$,
and which is impulsively kicked after a delay $\tau$. We derive the
expectation value $\braket{\hat{a}}$ only, because $\braket{\hat{a}^{\dagger}}=\braket{\hat{a}}^{*}$.
At the moment of the kick ($t=\tau$), the wave function reads
\begin{equation}
\ket{\psi_{-}}=e^{-|\alpha_{0}|^{2}/2}\sum_{n=0}^{\infty}\frac{\alpha_{0}^{n}}{\sqrt{n!}}e^{-iE_{n}\tau}\ket{n}.\label{eq:App-wf-before-kick}
\end{equation}
Using the completeness of coherent states, we expand each $\ket{n}$
state in terms of coherent states
\begin{equation}
\ket{n}=\frac{1}{\pi}\int_{\mathbb{C}}e^{-|\alpha|^{2}/2}\frac{(\alpha^{*})^{n}}{\sqrt{n!}}\ket{\alpha}\,d^{2}\alpha.\label{eq:App-number-state-expantion}
\end{equation}
In the impulsive approximation (for details see Subsection \ref{subsec:Mechanism-quantum-echo}),
the wave function immediately after the kick, reads
\begin{align}
\ket{\psi_{+}}\approx\hat{D}(\beta)\ket{\psi_{-}} & =\frac{e^{-|\alpha_{0}|^{2}/2}}{\pi}\times\nonumber \\
\sum_{n=0}^{\infty}\frac{\alpha_{0}^{n}}{n!}e^{-iE_{n}\tau} & \int_{\mathbb{C}}e^{-|\alpha|^{2}/2}(\alpha^{*})^{n}\hat{D}(\beta)\ket{\alpha}\,d^{2}\alpha,\label{eq:App-after-kick}
\end{align}
where $\hat{D}(\beta)=\exp(\beta\hat{a}^{\dagger}-\beta^{*}\hat{a})$
is the displacement operator, which shifts coherent states $\hat{D}(\beta)\ket{\alpha}=\exp[(\alpha^{*}\beta-\alpha\beta^{*})/2]\ket{\alpha+\beta}$.
Then, we express each coherent state $\ket{\alpha+\beta}$ in terms
of number states $\{\ket{m}\}$ and substitute $\beta=i\lambda$,
where $\lambda=-E_{0}\int_{-\infty}^{\infty}f(t)\,dt$. After rearrangement,
$\ket{\psi_{+}}$ reads
\begin{align}
\ket{\psi_{+}}=e^{-|\alpha_{0}|^{2}/2}\sum_{m,n=0}^{\infty}\frac{\alpha_{0}^{n}}{n!}\frac{1}{\sqrt{m!}}e^{-iE_{n}\tau}\ket{m} & \times\nonumber \\
\frac{1}{\pi}\int_{\mathbb{C}}e^{-|\alpha|^{2}}(\alpha^{*})^{n}e^{i\alpha\lambda-\lambda^{2}}(\alpha+i\lambda)^{m}\,d^{2}\alpha.\label{eq:App-wf-after-kick}
\end{align}

Keeping terms up to the second order in $\lambda$ in the complex
integral in Eq. (\ref{eq:App-wf-after-kick}), results in
\begin{align}
\frac{1}{\pi}\int_{\mathbb{C}}e^{-|\alpha|^{2}}(\alpha^{*})^{n}\left[\alpha^{m}-\lambda^{2}(1+m)\alpha^{m}\right.\nonumber \\
+i\lambda m\alpha^{m-1}+i\lambda\alpha^{m+1}\qquad\quad\nonumber \\
\left.+(\lambda^{2}/2)(m-m^{2})\alpha^{m-2}-(\lambda^{2}/2)\alpha^{m+2}\right] & \,d^{2}\alpha.\label{eq:App-integral-second-order}
\end{align}
The task now is to evaluate the six integrals in Eq. (\ref{eq:App-integral-second-order})
\begin{equation}
\begin{aligned}I_{f} & =\frac{1}{\pi}\int_{\mathbb{C}}e^{-|\alpha|^{2}}(\alpha^{*})^{n}\alpha^{m}\,d^{2}\alpha,\\
I_{j} & =\frac{C_{j}(m)}{\pi}\int_{\mathbb{C}}e^{-|\alpha|^{2}}(\alpha^{*})^{n}\alpha^{m+j}\,d^{2}\alpha,
\end{aligned}
\label{eq:App-small-integrals}
\end{equation}
where $j=-2,\dots,2$, and the coefficients are defined as
\begin{equation}
\begin{aligned}C_{-2}(m)=(\lambda^{2}/2)(m-m^{2}), & \quad C_{2}(m)=-(\lambda^{2}/2),\\
C_{-1}(m)=i\lambda m, & \quad C_{1}(m)=i\lambda,\\
C_{0}(m)=-\lambda^{2} & (1+m).
\end{aligned}
\label{eq:App-coefficients}
\end{equation}

To evaluate the integrals, we substitute $\alpha=re^{i\varphi}$,
change variables $t=r^{2}$, and use the definition of the Gamma function,
$\Gamma(n)=\int_{0}^{\infty}e^{-t}t^{2n}\,dt=(n-1)!$. This results
in $I_{f}=n!$ and $I_{j}=C_{j}(m-j)!$. Substituting these back into
Eq. (\ref{eq:App-wf-after-kick}) and propagating each number state
$\ket{m}$ by multiplying it with $\exp(-iE_{m}T)$, results in six
sums
\begin{equation}
\begin{aligned}\Sigma_{f} & =e^{-|\alpha_{0}|^{2}/2}\sum_{n=0}^{\infty}\frac{\alpha_{0}^{n}}{\sqrt{n!}}e^{-iE_{n}(\tau+T)}\ket{n},\\
\Sigma_{j} & =e^{-|\alpha_{0}|^{2}/2}\sum_{n=0}^{\infty}\frac{C_{j}\alpha_{0}^{n+j}}{\sqrt{n!}}e^{-i(E_{n+j}\tau+E_{n}T)}\ket{n}.
\end{aligned}
\label{eq:App-sums}
\end{equation}
\textbf{Note}: the time $T$ here is counted from the moment of the
application of the kick.

After acting with the operator $\hat{a}$, the first sum becomes
\begin{equation}
\hat{a}\Sigma_{f}=e^{-|\alpha_{0}|^{2}/2}\sum_{n=1}^{\infty}\frac{\alpha_{0}^{n}}{\sqrt{n!}}e^{-iE_{n}(\tau+T)}\sqrt{n}\ket{n-1},\label{eq:App-a-first-sum}
\end{equation}
and changing the index of summation from $n$ to $k=n-1$, results
in
\begin{equation}
\hat{a}\Sigma_{f}=e^{-|\alpha_{0}|^{2}/2}\sum_{k=0}^{\infty}\frac{\alpha_{0}^{k+1}}{\sqrt{k!}}e^{-iE_{k+1}(\tau+T)}\ket{k}.\label{eq:App-a-first-sum2}
\end{equation}
The rest of the sums become
\begin{align}
\hat{a}\Sigma_{j} & =e^{-|\alpha_{0}|^{2}/2}\sum_{n=1}^{\infty}\frac{C_{j}(n)\alpha_{0}^{n+j}}{\sqrt{(n-1)!}}\nonumber \\
 & \times e^{-i(E_{n+j}\tau+E_{n}T)}\ket{n-1},\label{eq:App-a-second-sum}
\end{align}
 and after changing the index of summation, the sums read
\begin{align}
\hat{a}\Sigma_{j} & =e^{-|\alpha_{0}|^{2}/2}\sum_{k=0}^{\infty}\frac{C_{j}(k+1)\alpha_{0}^{k+1+j}}{\sqrt{k!}}\nonumber \\
 & \times e^{-i(E_{k+1+j}\tau+E_{k+1}T)}\ket{k}.\label{eq:App-a-second-sum2}
\end{align}

The expectation value $\braket{\hat{a}}$ involves the products $(\Sigma_{f}^{*}+\Sigma_{j_{1}}^{*})(\hat{a}\Sigma_{f}+\hat{a}\Sigma_{j_{2}})$,
where $j_{1,2}=-2,\dots,2$. Without loss of generality, we assume
$\alpha_{0}$ is real, such that $\alpha^{*}=\alpha$ and $|\alpha|^{2}=\alpha^{2}$.
Considering the products second order in $\lambda$ and expanding
the exponentials using the expression for energy, the products read
\begin{equation}
\begin{aligned}\Sigma_{f}^{*}\hat{a}\Sigma_{f}= & \alpha_{0}e^{-|\alpha_{0}|^{2}}\\
\times\exp\{ & |\alpha_{0}|^{2}e^{-2i(T+\tau)}-i\Delta\cdot(T+\tau)\},\\
\Sigma_{f}^{*}\hat{a}\Sigma_{j}= & \alpha_{0}^{j+1}e^{-|\alpha_{0}|^{2}}\\
\times\exp[ & -i(j+\Delta\cdot j+j^{2})\tau-i\Delta\cdot(T+\tau)]\\
\times\sum_{k=0}^{\infty} & \frac{C_{j}(k+1)\left[\alpha_{0}^{2}e^{-2i(T+\tau+j\tau)}\right]^{k}}{k!}\\
 & j=-2,\dots,2,\\
\Sigma_{j}^{*}\hat{a}\Sigma_{f}= & \alpha_{0}^{j+1}e^{-|\alpha_{0}|^{2}}\\
\times\exp[-i & (j-\Delta\cdot j-j^{2})\tau-i\Delta\cdot(T+\tau)]\\
\times\sum_{k=0}^{\infty} & \frac{C_{j}^{*}(k)\left[\alpha_{0}^{2}e^{-2i(T+\tau-j\tau)}\right]^{k}}{k!}\\
 & j=-2,\dots,2,\\
\Sigma_{j_{1}}^{*}\hat{a}\Sigma_{j_{2}}= & e^{-|\alpha_{0}|^{2}}\alpha_{0}^{j_{1}+j_{2}+1}\\
\times & \exp[-i\tau(j_{1}+j_{2}+j_{2}^{2}-j_{1}^{2})]\\
\times & \exp\{-i\Delta\cdot[(j_{2}-j_{1})\tau+(T+\tau)]\}\\
\times\sum_{n=0}^{\infty}C_{j_{1}}^{*}(k) & C_{j_{2}}(k+1)\frac{\left[\alpha_{0}^{2}e^{-2i[T+\tau+(j_{2}-j_{1})\tau]}\right]^{k}}{k!}\\
 & j_{1,2}=\pm1.
\end{aligned}
\label{eq:App-expanding-exponentials}
\end{equation}
The first term $\Sigma_{f}^{*}\hat{a}\Sigma_{f}$ is independent of
$\lambda$, and it identical to Eq. (\ref{eq:App-free-expectation-a}),
describing the free propagation of a coherent state. The rest of the
terms are proportional to $\lambda,\lambda^{2}$ and arise due to
the kick. The series in Eq. (\ref{eq:App-expanding-exponentials})
are of the form $\sum_{k=0}^{\infty}f(k)z^{k}/k!$, where $f(k)$
(the $C_{j}$s) are polynomials in $k$. Such series can be summed
up, producing terms of the form $p(z)e^{z}$, where $p(z)$ is a polynomial
in $z$. The results are
\begin{equation}
\begin{aligned}\Sigma_{f}^{*}\hat{a}\Sigma_{j}= & \alpha_{0}^{j+1}e^{-|\alpha_{0}|^{2}}\\
\times & \exp[-i(j+j\Delta+j^{2})\tau-i\Delta\cdot(t+\tau)]\\
\times & g_{j}\left[\alpha_{0}^{2}e^{-2i(T+\tau+j\tau)}\right]\exp\left[\alpha_{0}^{2}e^{-2i(T+\tau+j\tau)}\right]\\
 & j=-2,\dots,2,\\
\Sigma_{j}^{*}\hat{a}\Sigma_{f}= & \alpha_{0}^{j+1}e^{-|\alpha_{0}|^{2}}\\
\times & \exp[-i(j-j\cdot\Delta-j^{2})\tau-i\Delta\cdot(t+\tau)]\\
\times & h_{j}\left[\alpha_{0}^{2}e^{-2i(T+\tau-j\tau)}\right]\exp\left[\alpha_{0}^{2}e^{-2i(T+\tau-j\tau)}\right]\\
 & j=-2,\dots,2,\\
\Sigma_{j_{1}}^{*}\hat{a}\Sigma_{j_{2}} & =\alpha_{0}^{j_{1}+j_{2}+1}e^{-|\alpha_{0}|^{2}}e^{-i\Delta\cdot(T+\tau)}\\
 & \exp\{-i[j_{1}+j_{2}+j_{2}^{2}-j_{1}^{2}+\Delta\cdot(j_{2}-j_{1})]\tau\}\\
 & w_{j_{1}j_{2}}\exp\left[\alpha_{0}^{2}e^{-2i[T+\tau+(j_{2}-j_{1})\tau]}\right]\\
 & j_{1,2}=\pm1,
\end{aligned}
\label{eq:App-expanding-exponentials-2}
\end{equation}
where the polynomials $g_{j}$, $h_{j}$, and $w_{j_{1},j_{2}}$ are
given by
\begin{equation}
\begin{aligned}g_{-2}(z)=-\frac{\lambda^{2}}{2}(2z+z^{2}), & \qquad g_{2}(z)=-\frac{\lambda^{2}}{2},\\
g_{-1}(z)=i\lambda(z+1), & \qquad g_{1}(z)=i\lambda,\\
g_{0}(z)=- & \lambda^{2}(z+2),
\end{aligned}
\label{eq:App-g-function}
\end{equation}

\begin{equation}
\begin{aligned}h_{-2}(z)=-\frac{\lambda^{2}}{2}z^{2}, & \qquad h_{2}(z)=-\frac{\lambda^{2}}{2},\\
h_{-1}(z)=-i\lambda z, & \qquad h_{1}(z)=-i\lambda,\\
h_{0}(z)= & -\lambda^{2}(z+1),
\end{aligned}
\label{eq:App-h-function}
\end{equation}
and
\begin{equation}
\begin{aligned}w_{-1,-1}(z)=\lambda^{2}\left(z^{2}+2z\right), & \qquad w_{-1,1}(z)=\lambda^{2}z,\\
w_{1,-1}(z)=\lambda^{2}(z+1), & \qquad w_{1,1}(z)=\lambda^{2}.
\end{aligned}
\label{eq:App-w-function}
\end{equation}
To simplify the notations, we define
\begin{equation}
\begin{aligned}A_{j} & =\alpha_{0}^{j+1}e^{-|\alpha_{0}|^{2}}\\
 & \times\exp\{-i[j(1+\Delta)+j^{2}]\tau-i\Delta\cdot t\},\\
B_{j}(t) & =\alpha_{0}^{j+1}e^{-|\alpha_{0}|^{2}}\\
 & \times\exp\{-i[j(1-\Delta)-j^{2}]\tau-i\Delta\cdot t\},\\
C_{j_{1},j_{2}}(t) & =\alpha_{0}^{j_{1}+j_{2}+1}e^{-|\alpha_{0}|^{2}}e^{-i\Delta\cdot t}\\
\times\exp\{-i & [j_{1}+j_{2}+j_{2}^{2}-j_{1}^{2}+\Delta\cdot(j_{2}-j_{1})]\tau\},
\end{aligned}
\label{eq:App-ABC-coefficients}
\end{equation}

\noindent where $z(t,j;\tau)=\alpha_{0}^{2}e^{-2i(T+j\tau)}$, and
$t$ counts from the beginning of the evolution. Finally, the expectation
value of the operator $\hat{q}=(\hat{a}^{\dagger}+\hat{a})/\sqrt{2}$
reads
\begin{equation}
\begin{aligned}\braket{\hat{q}}(t) & =[c_{f}+\lambda^{2}c_{0}(t)]e^{z(t)}+\lambda\sum_{j=\pm1}c_{j}(t)e^{z(t+j\tau)}\\
 & +\lambda^{2}\sum_{j=\pm2}c_{j}(t)e^{z(t+j\tau)}+\mathrm{c.c.},
\end{aligned}
\label{eq:App-q-after-kick-lambda2}
\end{equation}
where
\begin{equation}
\begin{aligned}c_{f} & =\frac{A_{0}}{\sqrt{2}},\\
\lambda^{2}c_{0}(t) & =\frac{1}{\sqrt{2}}\left(A_{0}g_{0}+B_{0}h_{0}\right.+\\
 & \left.+C_{11}w_{11}+C_{-1,-1}w_{-1,-1}\right),\\
\lambda^{2}c_{-2}(t) & =\frac{1}{\sqrt{2}}(A_{-2}g_{-2}+B_{2}h_{2}+C_{1,-1}w_{1,-1}),\\
\lambda^{2}c_{2}(t) & =\frac{1}{\sqrt{2}}(A_{2}g_{2}+B_{-2}h_{-2}+C_{-1,1}w_{-1,1}),\\
\lambda c_{-1}(t) & =\frac{1}{\sqrt{2}}(A_{-1}g_{-1}+B_{1}h_{1}),\\
\lambda c_{1}(t) & =\frac{1}{\sqrt{2}}(A_{1}g_{1}+B_{-1}h_{-1}).
\end{aligned}
\label{eq:App-c-coefficients}
\end{equation}

\begin{figure}[H]
\begin{centering}
\includegraphics[width=0.95\columnwidth]{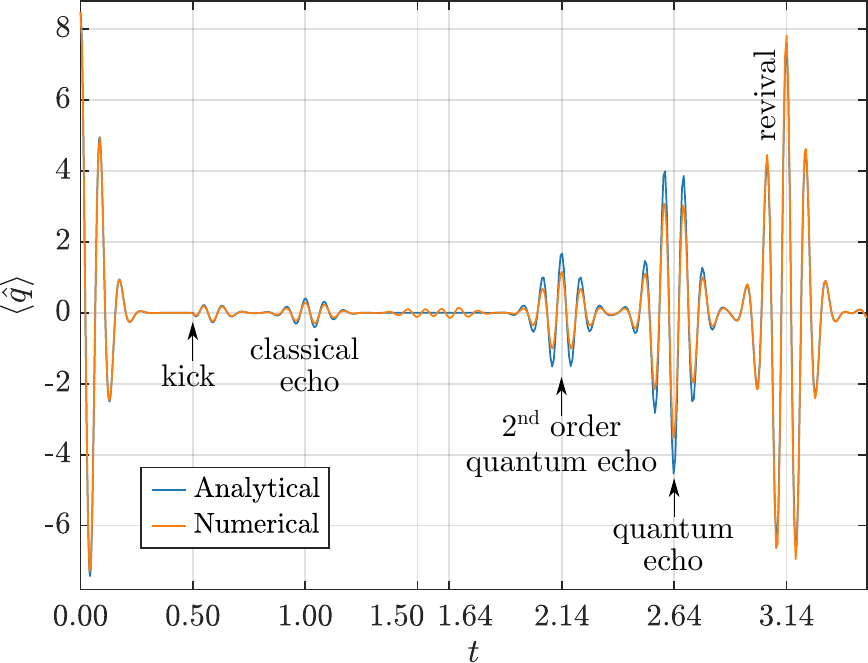}
\par\end{centering}
\caption{The quantum expectation value $\braket{\hat{q}}(t)$ obtained (i)
using Eqs. (\ref{eq:App-q-after-kick-lambda2}) and (\ref{eq:App-c-coefficients}),
and (ii) numerically solving the time-dependent Shr\"odinger equation
with the Hamiltonian in Eq. (\ref{eq:H-dimensionless}). The detuning
parameter is $\Delta=0.01$. Initial coherent state is defined by
$\alpha_{0}=6$. The kick is applied at $t=\tau=0.5$. Excitation
is Gaussian in time, $E_{0}f(t)=E_{0}\exp[-(t-\tau)^{2}/\sigma^{2}]$,
with $E_{0}=3.0$ and $\sigma=0.01$. The parameters are identical
to Fig. \ref{fig:echo-classical}. \label{fig:App-second-order-lambda}}
\end{figure}

Figure \ref{fig:App-second-order-lambda} compares the expectation
value $\braket{\hat{q}}(t)$ obtained using Eqs. (\ref{eq:App-q-after-kick-lambda2})
and (\ref{eq:App-c-coefficients}), with the results of direct numerical
integration of the time-dependent Shr\"odinger equation. The numerical
result shows the overlapping classical echo of the second order and
quantum echo of the third order between $t\approx1.50$ and $t\approx1.64$.
These echoes are higher order effects that emerge in the third order
perturbation theory in the parameter $\lambda$.

\end{document}